\documentstyle[twoside,fleqn,espcrc2,psfig]{article}

\newcommand{\AmS}{{\protect\the\textfont2
  A\kern-.1667em\lower.5ex\hbox{M}\kern-.125emS}}

\title{$d$-wave pairing symmetry in cuprate superconductors}

\author{C.C. Tsuei and J.R. Kirtley \address{IBM T.J. Watson Research Center, Yorktown
Heights, NY 10598, USA}}

\begin{document}

\begin{abstract}
Phase-sensitive tests of pairing symmetry have provided strong evidence
for predominantly $d$-wave pairing symmetry in both the hole- and
electron-doped high-T$_c$ cuprate superconductors. Temperature dependent
measurements in YBa$_2$Cu$_3$O$_{7-\delta}$ (YBCO)
indicate that the $d$-wave pairing dominates, with little
if any imaginary component, at all temperatures from 0.5K through T$_c$.
In this article we review some of this evidence and discuss
the implications of the universal $d$-wave pairing symmetry in the cuprates.
\end{abstract}

\maketitle

\section{INTRODUCTION}

Pairing symmetry in the cuprate superconductors was a subject of intense
debate for many years after the discovery of high-temperature
superconductivity \cite{bednorz}, primarily because the interpretation
of many conventional techniques, such as quasiparticle tunneling,
NMR, angle-resolved photoemission, and penetration depth measurements,
was model-dependent. Nevertheless,
these phase-insensitive techniques have produced a large body of evidence
for $d$-wave pairing in the cuprates.
The recent development of phase-sensitive
pairing symmetry test \cite{wollman,triprl,brawner,mathai,wollman2,miller},
has yielded compelling evidence for predominantly $d$-wave
pairing symmetry in a number of optimally doped cuprates \cite{symrmp}.
A question naturally arises: How universal is the $d_{x^2-y^2}$ pairing
in cuprate superconductors?

There are numerous theoretical studies suggesting the stability of the
$d$-wave pair state depends on the details of band structure and the
pairing potential
\cite{carbotte,wheatley,koltenbah}. There has also been considerable theoretical
studies indicating \cite{laughlin,laughlin2,bahcall,salkola,balatsky,sigrist}
that a pure $d_{x^2-y^2}$ pair state is not stable against the formation
of time reversal symmetry breaking states such as $d_{x^2-y^2}+id_{xy}$
or $d_{x^2-y^2}+is$, at surfaces, interfaces, near impurities, or
below a certain characteristic temperature. On the experimental side,
Raman data on Bi$_2$Sr$_2$CaCu$_2$O$_{8+\delta}$ and
Tl$_2$Ba$_2$CuO$_{8+\delta}$ systems as a function of oxygen content
indicate that the order parameter has $d$-wave symmetry near optimal
doping and isotropic pairing in the over-doped regime \cite{kendziora}. There
are additional indirect experimental evidences for the fragility of
the pure $d$-wave state \cite{ma,krishana}.
The following will examine the universality issues
based mainly on the results of phase sensitive
pairing symmetry experiments. We will conclude with a brief discussion on
implications of $d$-wave superconductivity in the cuprates.

\section{PHASE-SENSITIVE PAIRING SYMMETRY TESTS}

The sign and magnitude of the pair tunneling critical current
I$_c$
across a Josephson junction made with at least one superconductor
with unconventional pairing symmetry depends sensitively on the gap
function symmetry and relative orientation of the junction electrodes.
The
ground state of a superconducting loop with an odd number of negative
critical currents (termed a ``frustrated" or ``$\pi$"
loop) is doubly degenerate and shows spontaneous magnetization
of one-half flux quantum ($\Phi_0/2=hc/4e=1.035\times10$-7$G-cm^2$),
provided that $L\mid I_c \mid >> \Phi_0$, where $L$ is the self-inductance
of the ring and $I_c$ is the critical current of the weakest junction
in the loop \cite{bulaevski,geshkenbein,sigrice}.
By varying the loop geometry, the presence or absence of the half-integer
flux quantum effect can be used for a definitive determination of the
pairing symmetry.

The tricrystal pairing symmetry tests use a ring consisting
of three crystals with controlled orientation (see Fig. 1) to define
the direction of the pair wavefunction.
\begin{figure}
\centerline{\psfig{figure=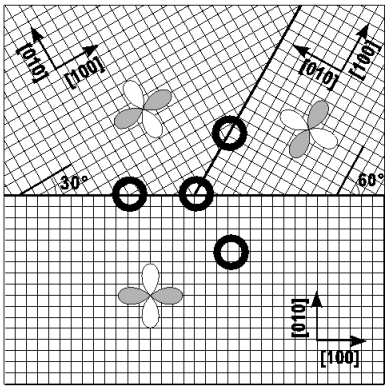,height=2in}}
\label{fig:trigeo}
\caption{A tricrystal geometrical configuration of the (100) SiTrO$_3$ (STO)
substrate designed for testing $d$-wave pairing symmetry in cuprate
superconductors. Also shown are polar plots indicating the orientation
of the $d_{x^2-y^2}$ order parameters aligned with the substrate
crystallographic axes; and the locations of the 0-, 2-, and 3-junction
rings.}
\end{figure}
The magnetic flux
threading through the tricrystal ring is measured with a scanning SQUID
microscope \cite{ssmapl} for different tricrystal configurations
to differentiate between various pairing symmetries. The $\Phi_0/2$
effect is intrinsic in a frustrated geometry, and is observed in ring,
disk, and blanket film samples at the tricrystal point.

\section{UNIVERSALITY of the $d$-WAVE PAIR STATE}

The establishment of $d$-wave pairing in some of
the cuprates has imposed well
defined constraints on possible models of high-temperature superconductivity,
but it does not specify the high-T$_c$ mechanism. To gain further insight
into the origin of high-temperature superconductivity, it is important
to determine whether the $d_{x^2-y^2}$ pairing symmetry is universal.
One can investigate the universality issues from the following aspects.

\subsection{Various cuprate systems}

Phase-sensitive SQUID interferometry \cite{wollman,brawner}, tricrystal
magnetometry \cite{triprl}, SQUID magnetometry \cite{mathai}, and
single junction interferometry \cite{wollman2,miller} experiments have
demonstrated $d$-wave pairing symmetry in YBCO single crystals or
$c$-axis oriented epitaxial films. In addition, tricrystal magnetometry
experiments have demonstrated $d$-wave pairing symmetry in various
high-T$_c$ hole-doped cuprate systems such as Tl$_2$Ba$_2$CuO$_{6+\delta}$,
GdBa$_2$Cu$_3$O$_{7-\delta}$, and Bi$_2$Sr$_2$CaCu$_2$O$_{8+\delta}$.
(For a more complete tabulation of the phase-sensitive experiments
see Ref. \cite{symrmp}). More recently, the pairing symmetry of the
electron-doped cuprate superconductors Nd$_{1.85}$Ce$_{0.15}$CuO$_{4-\delta}$
(NCCO) and Pr$_{1.85}$Ce$_{0.15}$CuO$_{4-\delta}$ (PCCO) has been
determined to also be predominantly $d$-wave
by observing the half-flux quantum
effect in $c$-axis thick blanket
films ($\sim$1$\mu$m thick) epitaxially grown on
tricrystal substrates with the configuration
of Fig. 1 \cite{nccoprl}.
Samples with two other geometries, designed to be unfrustrated
for a $d$-wave superconductor, do not show the half-flux quantum effect.
Shown in Fig. 2a is a scanning SQUID microscope image of a NCCO film
\begin{figure}
\centerline{\psfig{figure=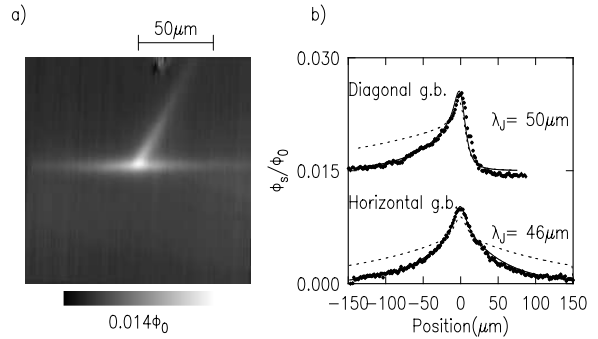,height=1.75in}}
\label{fig:nccodif}
\caption{(a) Scanning SQUID microscope image of the central area of a thick
film of NCCO epitaxially grown on a SrTiO$_3$ substrate with the
configuration depicted in Fig. 1. (b) Best fits to cross-sections through (a)
assuming the vortex has flux $\Phi=\Phi_0/2$ (solid line) and
$\Phi=\Phi_0$ (dashed line)}
\end{figure}
deposited on a frustrated tricrystal STO substrate, cooled to 4.2K in
nominal zero field. This image was obtained by subtracting an image with an externally
applied field of -0.15mG, which stabilized the half-quantum magnetic vortex
with the flux pointing down, from an image taken with an applied
field of 0.15mG, resulting in a half-quantum magnetic vortex pointing
up,
and dividing by 2. This results in an image of the half-quantum magnetic
vortex with effects from surface roughness,
a smoothly varying magnetic background, and the mutual inductance
between the SQUID and sample subtracted out. Figure 2b shows fitting of
cross-sections through the data of Fig. 2a
to expressions for the magnetic signals expected
from a Josephson vortex centered at the tricrystal point, with a total
flux of $\Phi_0/2$ \cite{blnktprl}.
The height $z$ of the pickup loop was determined by fitting a bulk
Abrikosov vortex.
The solid lines in
Fig. 2b are fits to cross-sectional
data for the horizontal and diagonal grain boundaries, assuming a total
flux of $\Phi_0/2$ for this vortex, using the
Josephson penetration depths $\lambda_J$ for the two grain boundaries
as the sole fitting parameters. The dashed lines, assuming a flux of
$\Phi_0$, are much worse fits. These results indicate that
the electron-doped superconductors, just as the hole-doped
superconductors, have $d$-wave pairing symmetry.

\subsection{Temperature dependence of $\Phi_0/2$}

It is of interest to find out whether $d$-wave pairing will change to
$s$-wave, for example, as a function of temperature. However, the
phase-sensitive experiments described above were all performed at liquid
helium temperature. With a recently developed variable sample temperature
scanning SQUID microscope \cite{vartapl},
the temperature dependence of the half-integer flux quantum effect in
tricrystal YBCO was studied. It was found that the total flux of the
Josephson vortex at the tricrystal point remains constant
(i.e. $\Phi_0/2$) from 0.5K through T$_c$ ($\sim$90K) \cite{vartsci}.
This finding means that the $d$-wave pair state dominates at all
temperatures below T$_c$, and that there is little, if any, time-reversal
symmetry breaking component in the order parameter, over the entire
temperature range. This is consistent with earlier studies, which found
less than a few percent of time-reversal symmetry breaking component
in phase-sensitive measurements at low temperatures \cite{spinnat,mathai}.

\subsection{The effect of doping}

Is there any band structure (doping) effect on pairing symmetry? In the
competition between $s$-wave and $d$-wave channels for high-temperature
superconductivity, is there any factor that can stabilize the $d$-wave
pairing? In the literature, there have been some theoretical studies
dealing with these issues: for example, based on a next-near-neighbor
pairing interaction, the symmetry of a BCS condensate is predicted to
vary as a function of energy band filling and other band parameters
\cite{carbotte,wheatley,koltenbah}. The $d$-wave pairing channel is
energetically favored in a wide range of doping centered around
half-filling. The stability of the $d$-wave pair state may be enhanced
by the proximity of the Fermi level to the van Hove singularity
in the 2D bands of the CuO$_2$ plane \cite{levin}.
A series of phase-sensitive
pairing symmetry tests as a function of doping in a model system
such as Tl$_2$Ba$_2$CuO$_{6+\delta}$, HgBa$_2$CuO$_{4+\delta}$,
Bi$_2$Sr$_2$CaCu$_2$O$_{8+\delta}$, or
Y$_{1-x}$Ca$_x$Ba$_2$Cu$_3$O$_{7-\delta}$ could test for these effects.

\section{CONCLUDING REMARKS}

Phase-sensitive experiments have produced strong
evidence for predominantly $d$-wave pairing symmetry in hole- and electron-doped
cuprate superconductors. These tests also indicate,
within the experimental accuracy ($\sim 3\%$ at low temperatures), the
absence of pairing with broken time-reversal symmetry. The universality
of $d$-wave pairing symmetry in bulk high-temperature superconductors is
thus well-established. The possibility of pair states without time-reversal
symmetry invariance (e.g. $d_{x^2-y^2}+id_{xy}$ or $d_{x^2-y^2}+is$)
induced at surfaces, interfaces, or around impurities or defects is
being actively investigated (see Ref. \cite{symrmp}
and references therein). Recent observations of a splitting of the
zero-bias conductance peak in quasiparticle tunneling \cite{covington}
and spontaneous magnetization in $c$-axis YBCO thin films
\cite{polturak,tafprl} represent possible evidence for such pair states.

Dominant $d$-wave pairing symmetry in both the hole-
and electron-doped cuprates is expected from single-band models of
the CuO$_2$ planes, in which the electron-hole symmetry is an
intrinsic property (see e.g. Poole {\it et al.} \cite{poole}).
From another perspective, the predominance of
$d_{x^2-y^2}$ pairing in cuprates underscores the important role of
strong electron correlation in determining the superconducting gap
symmetry and other properties. It is experimentally and theoretically
well established that a strong on-site Coulomb repulsion is
present in all cuprates. Such a strong electron correlation causes
the universally-seen Mott transition at half-filling. The strong
on-site Coulomb repulsion rules out simple $s$-wave pairing.

In future, a systematic study of pairing symmetry as
a function of doping, impurities, ... is important for a better
understanding of high-temperature superconductivity.

The authors wish to thank M. Bhushan, C.C. Chi, A. Gupta, Z.Z. Li, M.B. Ketchen, K.A. Moler,
D.M. Newns, H. Raffy, Z.F. Ren, J.Z. Sun, G. Trafas, and
J.H. Wang for invaluable assistance in the course of our
tricrystal experiments.

\end{document}